\documentclass{ws-ijmpc}
\usepackage[caption=false]{subfig}

\newcommand{\augx}{\bm \Phi_{\bm P(\bm X)}^{k,s}}
\newcommand{\augy}{\bm \Phi_{\bm P(\bm Y)}^{k,s}}

\begin{document}

\markboth{P. Fiedor \& A. Ho\l{}da}
{Time Evolution of Non-linear Currency Networks}

\catchline{}{}{}{}{}

\title{TIME EVOLUTION OF NON-LINEAR CURRENCY NETWORKS}

\author{PAWE\L{} FIEDOR}

\address{Cracow University of Economics\\
Rakowicka 27, 31-510 Krak\'{o}w, Poland\\
Pawel.F.Fiedor@ieee.org}

\author{ARTUR HO\L{}DA}

\address{Cracow University of Economics\\
Rakowicka 27, 31-510 Krak\'{o}w, Poland\\
aholda@uek.krakow.pl}

\maketitle

\begin{history}
\received{Day Month Year}
\revised{Day Month Year}
\end{history}

\begin{abstract}
Financial markets are complex adaptive systems, and are commonly studied as complex networks. Most of such studies fall short in two respects: they do not account for non-linearity of the studied relationships, and they create one network for the whole studied time series, providing an average picture of a very long, economically non-homogeneous, period. In this study we look at the currency markets by creating networks which can account for non-linearity in the underlying relationships, and are based on short time horizons with the use of running window approach. Since information--theoretic measures are slow to converge, we use Hirschfeld--Gebelein--R\'enyi Maximum Correlation Coefficient as a measure of the relationships between currencies. We use the Randomized Dependence Coefficient (RDC) as an estimator of the above. It measures the dependence between random samples as the largest canonical correlation between k randomly chosen non-linear projections of their copula transformations. On this basis we create full graphs, and further filter them into minimally spanning trees. We create such networks for each window moving along the studied time series, and analyse the time evolution of various network characteristics, particularly the degree distributions, and their economic significance. We apply this procedure to a dataset describing logarithmic changes in exchange rates in relation to silver for 27 world currencies for the years between 2002 and 2013.
\end{abstract}

\ccode{PACS Nos.: 05.10.-a, 64.60.aq, 89.65.-s}

\section{Introduction}

Financial markets are complex adaptive systems, even though the methodology of orthodox economics does not reflect this fact. Network theory, among other tools, is helpful in analysing the complexity of the interdependencies of economic entities, and particularly various financial instruments \cite{Mandelbrot:1963,Kadanoff:1971,Mantegna:1991}. Such analysis can lead to classifying the financial instruments or various risk assessments. The latter is particularly important in the analysis of currency markets. Econophysics developed a popular method of network analysis based on single linkage clustering analysis. What is interesting, most researchers are persistently using Pearson's correlation coefficient as the similarity measure in such analyses. Correlation structures have been analysed for time series describing stock returns \cite{Laloux:1999,Plerou:1999,Mantegna:1999,Tumminello:2010,Akemann:2011}, market index returns \cite{Bonanno:2000,Maslov:2001,Drozdz:2001,Coelho:2007,Gilmore:2008,Eryigit:2009,Song:2011,Sandoval:2012} and currency exchange rates \cite{McDonald:2005}. Since there is strong evidence of non-linear dynamics in stock returns \cite{Hsieh:1991,Brock:1991,Qi:1999,McMillan:2001,Sornette:2002,Kim:2002}, market index returns \cite{Franses:1996,Wong:1995,Chen:1996,Wong:1997,Scheinkman:1989,Ammermann:2003}, and currency exchange rate changes \cite{Hsieh:1989,Brock:1991,Rose:1991,Brooks:1996,Wu:2003}, we have introduced information--theoretic measures (mutual information and mutual information rate) as measures of similarity which can take non-linear dependencies into account \cite{Fiedor:2014a}.

Most analyses of this kind concentrate on the static structure of the studied markets, estimating the similarity measure for the whole available time series. There are at least two problems with this approach. First, particularly in the case of studying daily data, such analyses usually span across years or decades. In these cases the studied time series do not contain economically homogeneous data. This raises questions about the methodological propriety of averaging the analysis and results across such varied data. Second, even assuming this is methodologically sound, it only works well as an example in literature, but gives virtually no interesting insight for market participants, who are interested in the current state of the market, and the current trends, as opposed to an average state of the studied market in the last decade. Particularly the risk analyses need to be focused on short time horizons, as risk is inherently bound to the present. Relatively few inquiries look into these networks at short time horizons, and their time evolutions. Such analyses tend to be focused on the critical points and phase transitions \cite{Albert:2000,Dorogovtsev:2008,Sienkiewicz:2013,Wilinski:2013,Kozlowska:2014}. The evolution of various characteristics in networks have not be studied sufficiently, particularly in the case of currency markets \cite{Mizuno:2006,Ortega:2006,Naylor:2007,Kocakaplan:2012}. In this study we will concentrate on such issues, and we will study the currency market, as the time evolution of this market is particularly interesting for a wide range of economic entities, because currency rates affect all businesses which operate on a multinational level.

In this study we want to look at the currency markets by creating networks which can account for non-linearity of the underlying relationships, and are based on short time horizons with the use of running window approach. This will allow us to study the time evolution of these networks, their characteristics, and the currency market itself. In our previous studies we have used information--theoretic measures, such as mutual information, to account for non-linear behaviour in the financial markets. Mutual information equals zero if and only if the two studied random variables are strictly (statistically) independent. The estimation of mutual information in dynamical systems faces some difficulties however. These need to be understood in practical applications, but are usually not severe \cite{Paninski:2003,Palus:2001,Floros:2007,Hsieh:2009,Manap:2011,Steuer:2002,Papana:2009}. In our particular case we do run into one of these problems, namely information--theoretic measures are famously slow to converge, and we want to use narrow window in this study. For this reason we do not want to use mutual information, but we cannot fall back to using Pearson's correlation either, as it is a strictly linear measure. To include non-linearity, we use Hirschfeld--Gebelein--R\'enyi Maximum Correlation Coefficient as a measure of the relationships between currencies. As it is a theoretical concept, we use the Randomized Dependence Coefficient (RDC) as an estimator for practical purposes. The RDC measures the dependence between random samples as the largest canonical correlation between $k$ randomly chosen non-linear projections of their copula transformations.

There is another purpose of this study stemming from the use of the Randomized Dependence Coefficient, that is showing its usefulness as a similarity measure in the analysis of financial (and other) networks. Most studies of financial networks still use Pearson's correlation coefficient. This study may help researchers adapt the inclusion of non-linear relationships within financial data in their analyses. Part of the reason why researchers do not do so already may rest with the fact that information--theoretic measures are less known, and that they require discrete data. The discretisation step is not particularly troubling, but it may be viewed as an unnatural part in the analysis by some. Thus the use of the RDC may help promote the inclusion of non-linear relationships in network analyses of financial markets. Permutation entropy \cite{Bandt:2002} may also be used for such purpose, but we believe that it does not work as well as the above mentioned techniques in the case of financial data.

To study the world currency market we create these networks for a dataset describing daily logarithmic changes in exchange rates of 27 major world currencies for the period between January of 2002 and December of 2013. This period is long and varied enough to provide this study with interesting material to analyse. The study of currency markets requires one additional choice which is not necessary in the study of stock markets. Currency exchange rates need to be expressed in terms of base currency. It is standard to denominate currencies in terms of the major currency, usually the dollar. In financial network analyses this would mean excluding the dollar from the network however, which is not desirable. Two approaches can be used in this case, depending on the scope of the study. First, if a risk analysis for a company whose operations are based on a given currency is to be performed, the exchange rates should be denominated in terms of this (home) currency. Otherwise, as is the case in this paper, another base currency is used, most often the price of an ounce of gold, silver, or platinum. We have used silver for the purpose of this study, as gold is seen as less stable and more prone to speculation, thus not as effective for this purpose.

This paper is organised as follows. In Section 2 we present the methods used in the analysis, together with the rationale for the particular parameters used. In Section 3 we present the dataset used in this study and the obtained results. In Section 4 we discuss the mentioned results, and their economic significance. In Section 5 we conclude the study and propose further research.

\section{Methods}

In this section we present the methodology used in this study. In particular we present the Randomized Dependence Coefficient, and how we use it to create full graphs and then filter them into their minimally spanning trees.

The measuring of statistical dependence between random variables is a problem of enormous importance in various fields using statistical methods, particularly so in complex systems. Commonly used measures of dependence, among them principally Pearson's rho, Spearman's rank and Kendall's tau, are computationally efficient and their characteristics are well understood, but they deal only with a limited class of association patterns, namely linear or monotonically increasing functions. This is obviously problematic in analysing complex systems with various intricate dependencies. Proposing non-linear dependence measures is challenging however. This stems mostly from the sheer amount of possible association patterns. Nonetheless, many non-linear statistical dependence measures have indeed been proposed, including the Alternating Conditional Expectations \cite{Breiman85,Hastie86}, Kernel Canonical Correlation Analysis \cite{Bach02}, Copula Maximum Mean Discrepancy \cite{Gretton05,Gretton12,Poczos12}, Distance or Brownian Correlation \cite{Szekely07,Szekely10} and the Maximal Information Coefficient \cite{Reshef11}. Note that here we ignore the information--theoretic measures mentioned above, as they are a separate class, and we are not interested in them in this study. These methods are computationally expensive, show poor performance where additive noise is present, or are limited to scalar random variables. In this study we use the recently proposed Randomized Dependence Coefficient (RDC) \cite{Lopez:2013}, an estimator of the Hirschfeld-Gebelein-R\'enyi Maximum Correlation Coefficient (HGR) addressing these issues. RDC defines dependence between two random variables as the largest canonical correlation between random non-linear projections of their respective empirical copula-transformations. RDC is therefore invariant to monotonically increasing transformations, can operate on random variables of arbitrary dimension, and has a reasonable computational cost of $O(n\log n )$ with respect to the sample size.

In 1959 \cite{Renyi59}, A. R\'enyi proposed seven properties that a measure of statistical dependence $\rho^* : \mathcal{X} \times \mathcal{Y} \rightarrow [0,1]$ between random variables $X\in\mathcal{X}$ and $Y\in\mathcal{Y}$ should satisfy:
\begin{enumerate}
  \item $\rho^*(X,Y)$ is defined for any pair of non-constant random variables $X$ and $Y$.
  \item $\rho^*(X,Y) = \rho^*(Y,X)$
  \item $0 \leq \rho^*(X,Y) \leq 1$
  \item $\rho^*(X,Y) = 0$ iff $X$ and $Y$ are statistically independent.
  \item For bijective Borel-measurable functions $f,g : \mathbb{R} \rightarrow \mathbb{R}$, $\rho^*(X,Y) = \rho^*(f(X),g(Y))$.
  \item $\rho^*(X,Y) = 1$ if for Borel-measurable functions $f$ or $g$, $Y = f(X)$ or $X = g(Y)$.
  \item If $(X,Y) \sim \mathcal{N}(\bm \mu, \bm \Sigma)$, then $\rho^*(X,Y) = |\rho(X,Y)|$, where $\rho$ is the correlation coefficient.
\end{enumerate}
R\'enyi showed that the Hirschfeld-Gebelein-R\'enyi Maximum Correlation Coefficient \cite{Gebelein41,Renyi59} satisfies all these properties. It was defined by Gebelein \cite{Gebelein41} as the supremum of Pearson's linear correlation coefficient $\rho$ over all Borel-measurable functions $f,g$ of finite variance:
\begin{equation}
  \text{hgr}(X,Y) \equiv \sup_{f,g} \rho(f(X),g(Y)),
\end{equation}
Since the supremum is over an infinite-dimensional space, this measure is not directly computable. The Randomized Dependence Coefficient is a scalable estimator with the same structure as the Hirschfeld-Gebelein-R\'enyi Maximum Correlation Coefficient \cite{Lopez:2013}.

Given two random samples $\bm X \in \mathbb{R}^{p\times n}$ and $\bm Y \in \mathbb{R}^{q\times n}$ together with the parameters $k \in \mathbb{N}_+$ and $s \in \mathbb{R}_+$, the Randomized Dependence Coefficient between $\bm X$ and $\bm Y$ is defined as\cite{Lopez:2013}:
\begin{equation}
  \text{rdc}(\bm X, \bm Y; k,s) :=
  \sup_{\bm \alpha, \bm \beta}\rho\left(\bm \alpha^T \augx, \bm \beta^T
  \augy\right).
\end{equation}

From this we find that the calculation of the RDC follows three steps. We present them briefly, step by step. In order to achieve invariance with respect to transformations on marginal distributions, the RDC operates on the empirical copula transformation of the data \cite{Nelsen06,Poczos12}. If we consider a random vector $\bm X = (X_1, \ldots, X_d)$ with continuous marginal cumulative distribution functions $P_i$, $1 \leq i \leq d$, then the vector $\bm U = (U_1,\ldots,U_d) := \bm P(\bm X) = (P_1(X_1),\ldots,P_d(X_d))$ is known as the copula transformation, and has uniform marginals.

In the next step, the empirical copula transformations defined above are augmented with non-linear projections, in order for us to be able to use linear methods to capture non-linear dependencies within the original data. This idea has been widely used. In particular, Rahimi \& Brecht proved that linear regression on random, non-linear projections of the original feature space can generate high-performance regressors \cite{Rahimi08}. The choice of the non-linear dependencies $\phi : \mathbb{R} \rightarrow \mathbb{R}$ used is the main assumption in calculating the RDC. This is a problem found in all non-linear regression methods, and cannot be avoided. The choice of the family of features, and of probability distributions over it, is unlimited. The only way to choose is to use prior assumptions about which kind of distributions the method will typically have to analyse in a given situation. Commonly sigmoids, parabolas, radial basis functions, complex sinusoids, sines and cosines are used. In our study, we follow the authors of this coefficient \cite{Lopez:2013} and use sine and cosine projections, $\phi(\bm w^T \bm x +b) := (\cos(\bm w^T \bm x + b), \sin(\bm w^T \bm x + b))$. With this chose, shift-invariant kernels are approximated with these features when using the appropriate random parameter sampling distribution \cite{Rahimi08}. Additionally, functions with absolutely integrable Fourier transforms are approximated with $L_2$ error below $O(1/\sqrt{k})$ by $k$ of these features \cite{Jones92}.

Let the random parameters be distributed as follows: $\bm w_i \sim \mathcal{N}(\bm 0, s\bm I)$, $b_i \sim \mathcal{U}[-\pi,\pi]$. Choosing $\bm w_i$ to be Normal is equivalent to the use of the Gaussian kernel for MMD, CMMD or KCCA \cite{Rahimi08}. Tuning $s$ is equivalent to selecting the kernel width, or to regularise the non-linearity of the random projections \cite{Lopez:2013}.

Given a dataset $\bm X = (\bm x_1, \ldots, \bm x_n)$, we denote by
\begin{equation}
  \bm \Phi(\bm X; k,s) := 
  \left(
  \begin{array}{ccc}
  \phi(\bm w_1^T \bm x_1+b_1) & \cdots & \phi(\bm w_k^T \bm x_1+b_k)\\
  \vdots & \vdots & \vdots \\
  \phi(\bm w_1^T \bm x_n+b_1) & \cdots & \phi(\bm w_k^T \bm x_n +b_k)
  \end{array}
  \right)^T
\end{equation}
the $k-$th order random non-linear projection from $\bm X \in \mathbb{R}^{d \times n}$ to $\bm \Phi^{k,s}_{\bm X} := \bm \Phi(\bm X; k, s) \in \mathbb{R}^{2k \times n}$.

The final, third step, consists in computing the linear combinations of the augmented empirical copula transformations which have maximal correlation. We use the Canonical Correlation Analysis \cite{Haerdle07} as the calculation of pairs of basis vectors $(\bm \alpha, \bm \beta)$ such that the projections $\bm \alpha^T \bm X$ and $\bm \beta^T \bm Y$ of two random samples $\bm X \in \mathbb{R}^{p\times n}$ and $\bm Y \in \mathbb{R}^{q\times n}$ are maximally correlated \cite{Lopez:2013}.

With regards to the parameter selection we follow the authors of the RDC, as our testing showed that their choice seems to give good results. The number of random features for RDC was set to $k=10$ for both random samples, as we have not seen any significant improvements for larger values. Nonetheless, this parameter can be set to the largest value that fits within the available computational power. We set the random sampling parameters $(s_{\bm X}, s_{\bm Y})$ independently for each of the two random samples, as equal to their squared euclidean distance empirical median \cite{Gretton12}.

After selecting the parameters we describe the random samples which will be fed into the RDC in this study. Let us denote the most recent exchange rate for currency $e$ occurring on the end of day $t$ during the studied period by $p_e(t)$. Then for each currency the logarithmic rate changes are sampled,
\begin{equation}
 r_{e,t} \equiv \log(p_e(t)) - \log(p_e(t-1)),
\end{equation}
throughout the studied period. These time series constitute columns in a matrix $\bm{R}$. From these matrices an empirical correlation matrix $\bm{C}$ is constructed using the Randomized Dependence Coefficient of columns of matrix $\bm{R}$,
\begin{equation}
\bm{C}_{f,e} = \text{rdc}(\bm{R}_f, \bm{R}_e; k,s).
\label{eqn:corr_matrix}
\end{equation}

From the correlation matrix $R$ we create a matrix of Euclidean distances between the studied currencies $D$ in the following way\cite{Mantegna:1997}:
\begin{equation}
\bm{D}_{f,e} = \sqrt{2(1-\bm{C}_{f,e})}.
\end{equation}

Having defined the distance measure based on the Randomized Dependence Coefficient we describe the construction methods for filtered graphs, which we use in this study. Hierarchical networks based on such distance matrices can be constructed in two ways: either by forcing topological restraints or setting a threshold on the similarity measure. We use the first approach to create minimal spanning trees (MST). Threshold networks are very robust with regards to the statistical uncertainty in the estimation of similarity measures, but it is difficult to find a single threshold to appropriately display the nested structure of the similarity matrix. The method we use creates intrinsically hierarchical networks, but this approach is less stable with respect to the statistical uncertainty in the data. The distance matrix $\bm{D}$ containing $\bm{D}_{f,e}$ for all studied pairs is used to determine the minimal spanning tree \cite{Papa:1982} connecting $N$ financial instruments in the studied set in the following way. On the basis of the distance matrix $\bm{D}$ we create an ordered list $\mathcal{S}$, which contains the distances, listed in decreasing order. Then, to create a minimal spanning tree, starting from the first element of the list, the corresponding link is added to the network if and only if the resulting graph is still a forest or a tree \cite{Tumminello:2005}. After all appropriate links are added the graph is guaranteed to reduce to a tree. For detailed description of these methods see presented references \cite{Tumminello:2005,Lillo:2010}.

\section{Results}

To study the world currency market with the presented methodology we need a database of currency exchange rates. For this study we have used currency exchange rates for 27 major world currencies (listed in the appendices) from the Sauder School of Business database\footnote{http://fx.sauder.ubc.ca/data.html}. This database is based on official daily rates used and issued by the Bank of Canada. We have used end of day currency rates between the 1st of January 2002 and the 31st of December 2013, giving us the sample size of 3200. We use currency rates denominated in XAG (ounce of silver). We consider log changes of these rates, as explained above. We take the full sample of 3200 observations and use a running window approach of length equal 100 days (roughly 3 months), and create minimally spanning trees for every window. We present the results for a given network as connected to the last day of the 100 day period, that is the results presented for a given day contain only information available on and before that day.

\begin{figure}%
\centering
\subfloat[][]{%
\label{fig:intra-a}%
\includegraphics[width=0.5\textwidth]{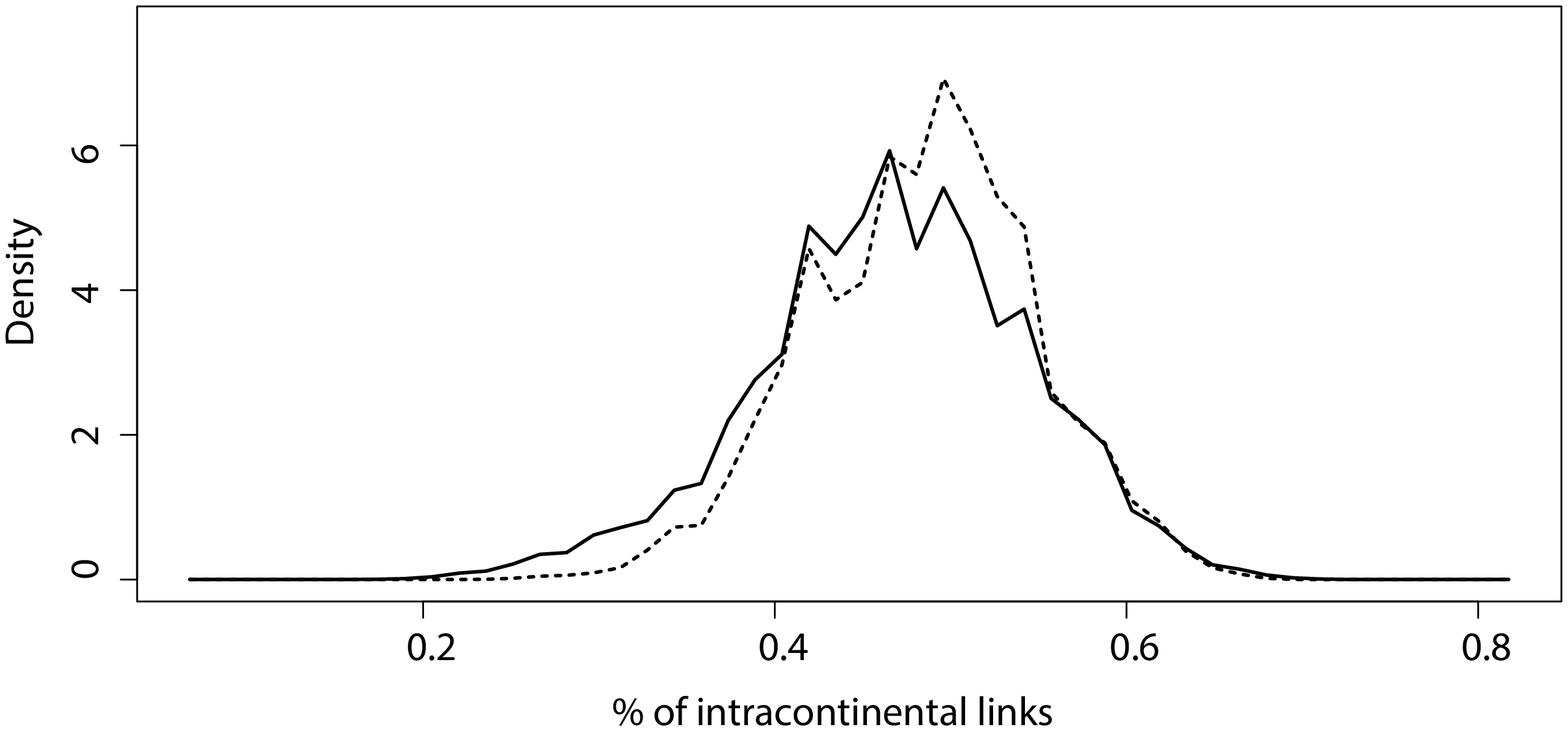}}%
\subfloat[][]{%
\label{fig:intra-b}%
\includegraphics[width=0.5\textwidth]{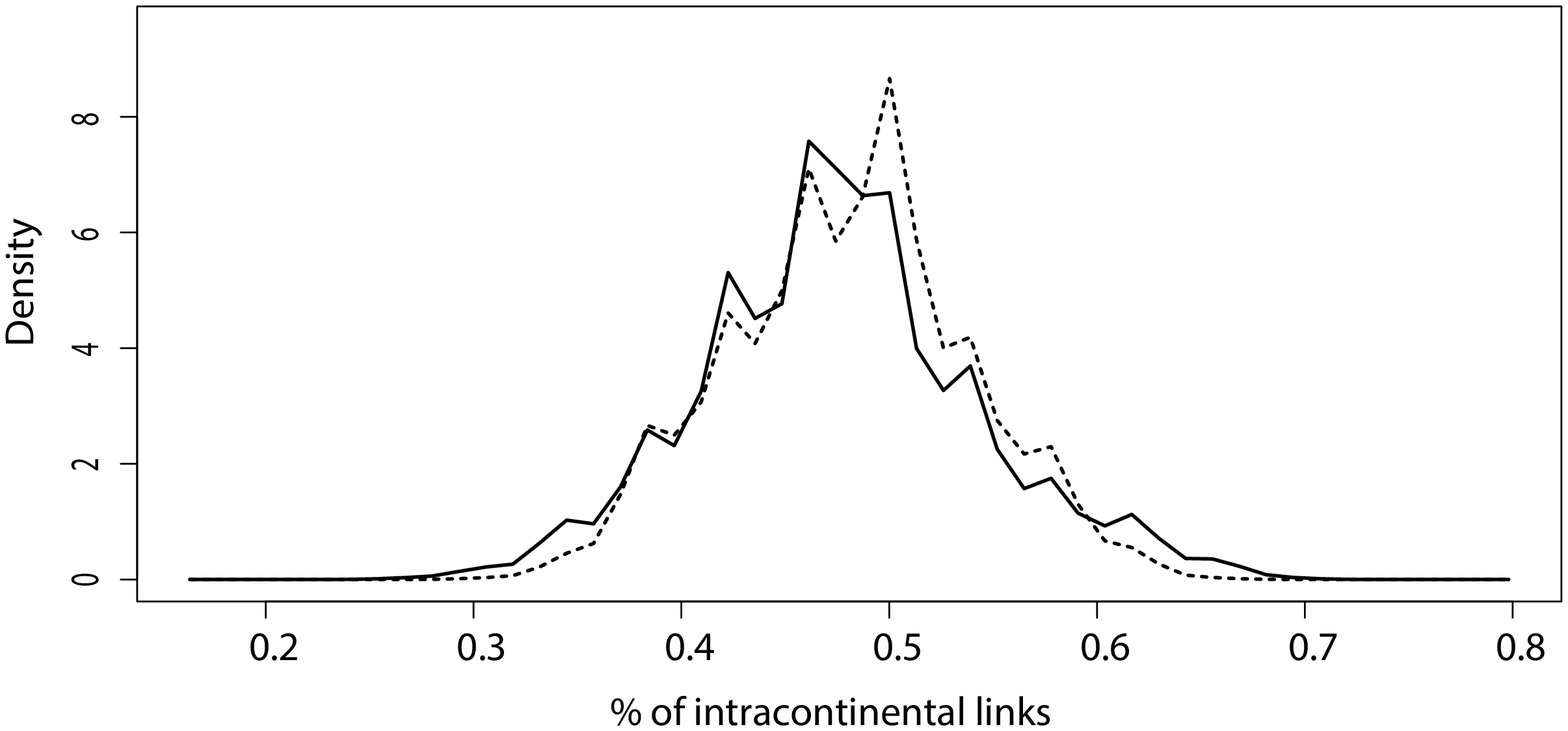}}%
\caption[]{Distribution of the percentage of intracontinental links in all links within all the calculated minimally spanning trees based on the Randomized Dependence Coefficient (solid lines) and the Pearson's correlation coefficient (dashed lines) for the various sizes of running window of:
\subref{fig:intra-a} 100 days;
\subref{fig:intra-b} 200 days.
It appears that the networks based on the Randomized Dependence Coefficient retain the ability to reconstruct the geographical structure of the currency market from the changes in currency rates themselves. This is an important feature of these networks and as such this result validates the methodological propriety of using the RDC in such analyses to include the non-linear dependencies.}%
\label{fig:intra}%
\end{figure}

First, we briefly show that this approach retains the valuable characteristics of the approach using Pearson's correlation coefficient, and then we analyse the time evolution of the obtained networks. Regarding the former, we investigate whether our method retrieves the geographical structure of the currencies from the price changes, which is ones of the important features of the standard methodology based on Pearson's correlation. In our previous studies we have shown this for networks based on mutual information. Having only one or a few networks we directly compared the percentage of intracontinental links in all links within the MST. Here we have 2920 networks for window of 100 days, thus we can investigate the distribution of the percentage of intracontinental links in all links within the MST and compare via kernel density estimation, which is presented in Fig.~\ref{fig:intra}.

Second, we turn to the analysis of the networks themselves. To start with, we present the overall picture of the studied period, that is the degree of all currencies, averaged over all created networks, to show the ranking of currencies' importance to the world market in the whole studied period. This is presented in Table~\ref{tab:ranks}.

\begin{table}[ht]
\tbl{Ranking of currencies in the whole studied period. The rank in parentheses denotes ranking based on Pearson's correlation coefficient as opposed to the main ranking where networks based on the Randomized Dependence Coefficient are used. Degree of all studied currencies, averaged over all networks created with the running window approach, is presented, and determines the rank. The analysis shows high importance of Asian currencies, even higher than the euro.}
{\begin{tabular}{llrllrllr}
\toprule
Rank & CCY & Av. D. & Rank & CCY & Av. D. & Rank & CCY & Av. D. \\ \colrule
1 (1) &  USD  &  4.321       & 10 (10) &  RUB  &  1.944       & 19 (15) &  TRY  &  1.438       \\
2 (2) &  HKD  &  3.829       & 11 (16) &  PKR  &  1.684       & 20 (21) &  NZD  &  1.433       \\
3 (4) &  CNY  &  3.141       & 12 (11) &  MXN  &  1.667       & 21 (23) &  ILS  &  1.387       \\
4 (6) &  EUR  &  2.621       & 13 (22) &  ARS  &  1.644       & 22 (18) &  CLP  &  1.381       \\
5 (3) &  DKK  &  2.602       & 14 (13) &  PHP  &  1.621       & 23 (17) &  NOK  &  1.373       \\
6 (5) &  TWD  &  2.578       & 15 (19) &  INR  &  1.599       & 24 (25) &  BRL  &  1.260       \\
7 (7) &  MYR  &  2.328       & 16 (14) &  GBP  &  1.525       & 25 (24) &  SEK  &  1.249       \\
8 (8) &  THB  &  1.969       & 17 (20) &  CHF  &  1.519       & 26 (26) &  ZAR  &  1.233       \\
9 (9) &  AUD  &  1.954       & 18 (12) &  CAD  &  1.496       & 27 (27) &  FJD  &  1.204       \\ \botrule
\end{tabular}
\label{tab:ranks}}
\end{table}

Then we begin to analyse the evolution of this market by presenting the three most important currencies (according to the degree distribution as above), but this time for every year in the studied period. This is presented in Table~\ref{tab:yearly}.

\begin{table}[ht]
\tbl{Ranking of the three main currencies in yearly sub-periods within the studied period based on their degree distributions, averaged over all networks created for a given year. The networks are based on the Randomized Dependence Coefficient. The analysis shows that the Euro lost its strong position around 2004, and have been losing it until 2010-2011, after which it is regaining it slowly. The dollar has lost its hegemony in the years of 2006 and 2009 to the renminbi, which may be related to the financial crisis. Additionally the Hong Kong dollar has become the leader of this ranking for the years of 2011 and 2013, hinting that the power of the dollar is not yet regained fully after the crisis. However in 2007 and 2008 the USD has been very strongly influencing the networks, which represents its role in the crisis.}
{\begin{tabular}{rllllll}
\toprule
Year &  Ranked \#1     &  Av. Degree  & Ranked \#2     &  Av. Degree   & Ranked \#3     &  Av. Degree  \\ \colrule
2002 & EUR &      4.138     & TWD &      3.938     & USD &      3.300     \\ 
2003 & USD &      4.425     & EUR &      3.468     & HKD &      3.302     \\ 
2004 & USD &      4.607     & HKD &      3.706     & DKK &      3.619     \\ 
2005 & USD &      4.372     & MYR &      3.888     & CNY &      3.516     \\ 
2006 & CNY &      4.908     & HKD &      4.640     & USD &      3.228     \\ 
2007 & USD &      5.996     & DKK &      3.462     & RUB &      3.147     \\ 
2008 & USD &      5.631     & HKD &      3.095     & DKK &      2.802     \\ 
2009 & CNY &      4.944     & TWD &      3.036     & HKD &      3.016     \\ 
2010 & USD &      4.490     & HKD &      4.351     & TWD &      3.514     \\ 
2011 & HKD &      4.752     & USD &      4.132     & CNY &      3.312     \\ 
2012 & USD &      4.713     & HKD &      4.641     & TWD &      3.908     \\ 
2013 & HKD &      4.804     & USD &      3.668     & TWD &      3.472     \\ \botrule
\end{tabular}
\label{tab:yearly}}
\end{table}

\begin{figure}%
\centering
\subfloat[][]{%
\label{fig:eurusd-a}%
\includegraphics[width=0.5\textwidth]{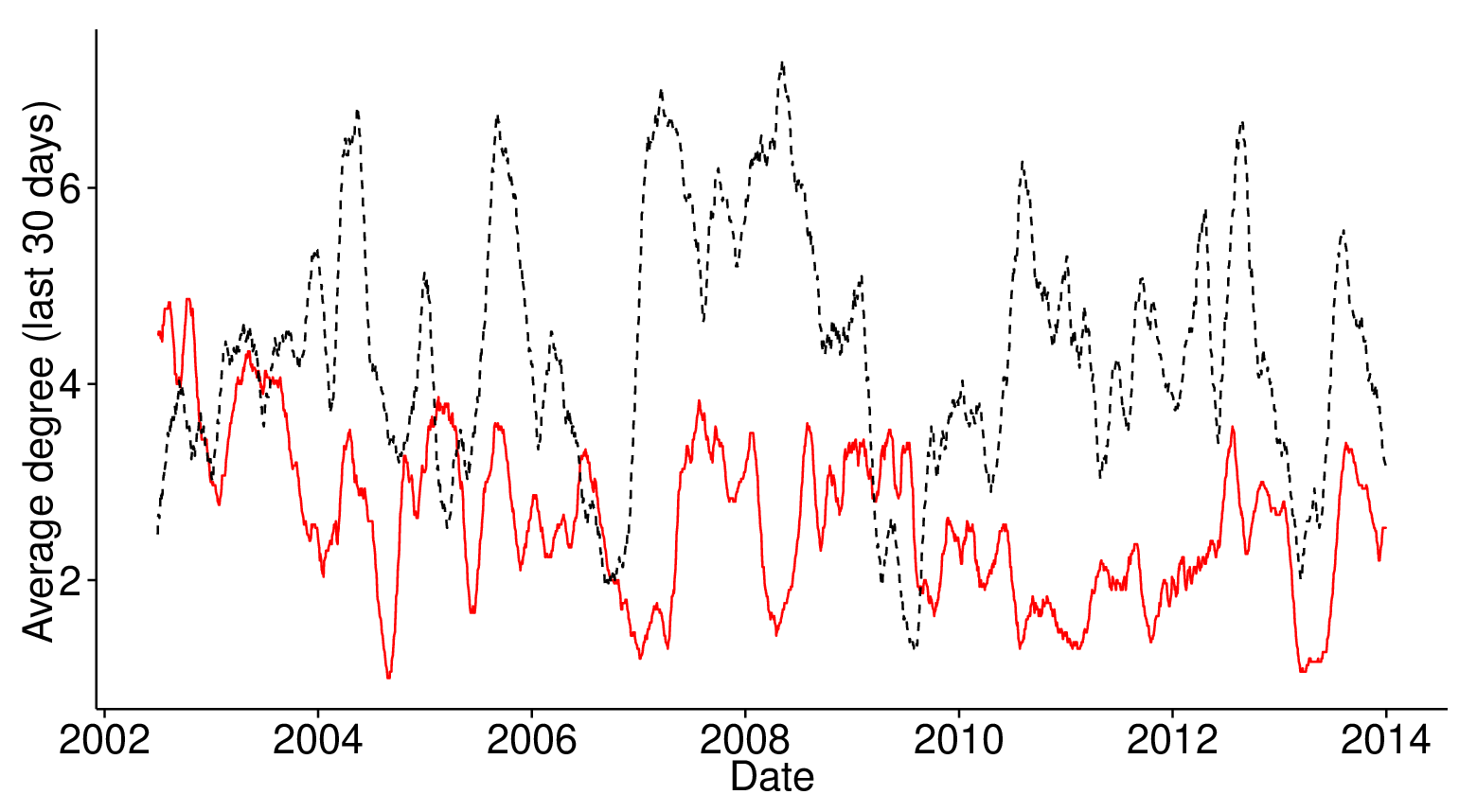}}%
\subfloat[][]{%
\label{fig:eurusd-b}%
\includegraphics[width=0.5\textwidth]{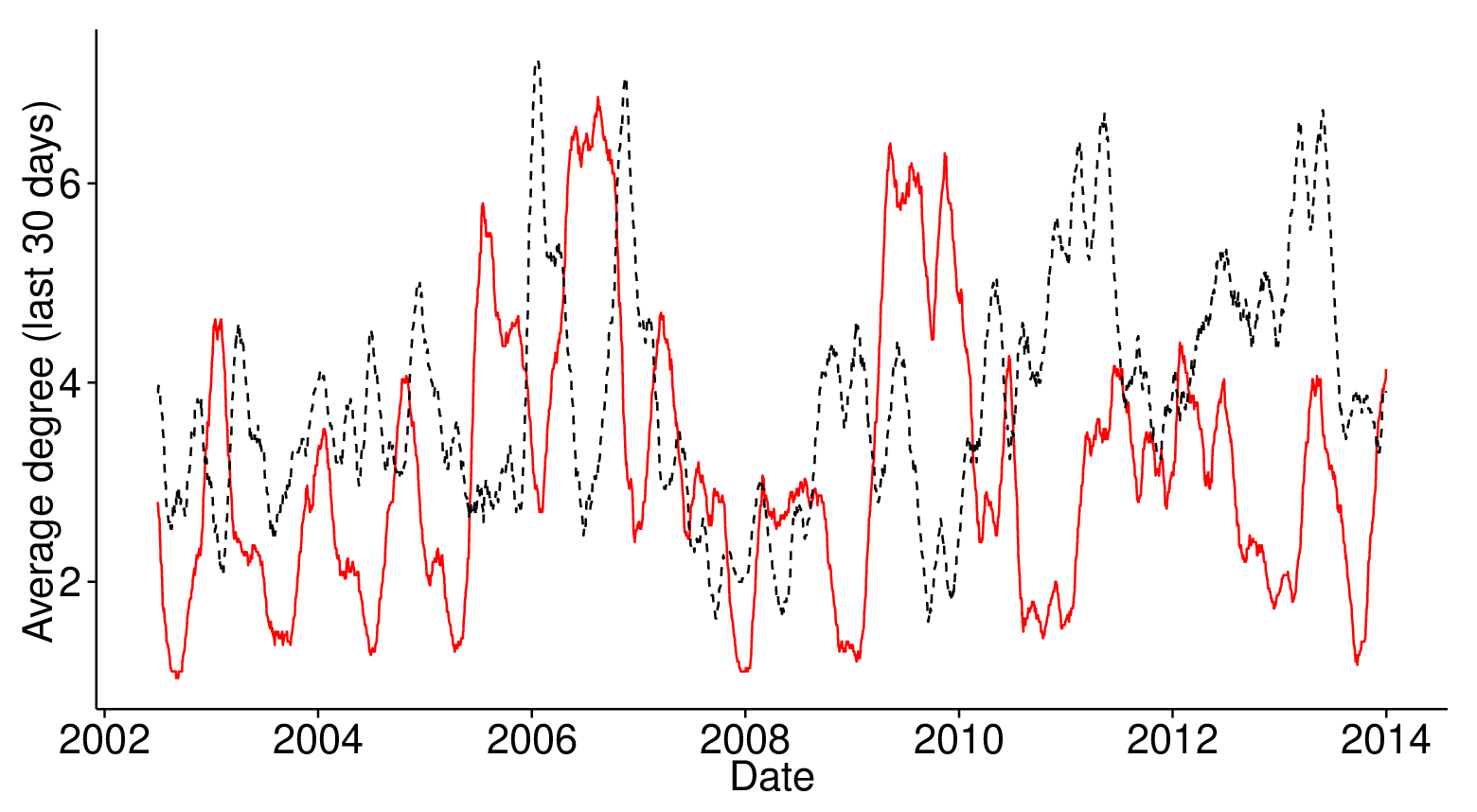}}%
\caption[]{Average degree of pairs of currencies for the last 30 networks calculated by a given date, based on minimally spanning trees based on the Randomized Dependence Coefficient, for:
\subref{fig:eurusd-a} EUR (solid) \& USD (dashed);
\subref{fig:eurusd-b} CNY (solid) \& HKD (dashed).
We see that while the EUR has been more important than the dollar in 2002, its power has dwindled. The dollar has been particularly strong around the time of the financial crisis of 2007-2008, at which time both CNY and HKD had their power diminished. There is also a negative correlation between degrees of Western and Asian currencies of $-0.337$ for CNY/USD, $-0.204$ for CNY/EUR, $-0.264$ for USD/HKD, and $-0.371$ for EUR/HKD. Correlation equals 0 for both USD/EUR and CNY/HKD pairs.}%
\label{fig:eurusd}%
\end{figure}

The case of the two main currencies in the West (EUR \& USD) and the two main currencies in the East (CNY \& HKD) are of particular interest, thus we present the evolution of their degree, averaged over 30 consecutive networks, in Fig.~\ref{fig:eurusd}.

\begin{figure}
\centerline{\includegraphics[width=0.9\textwidth]{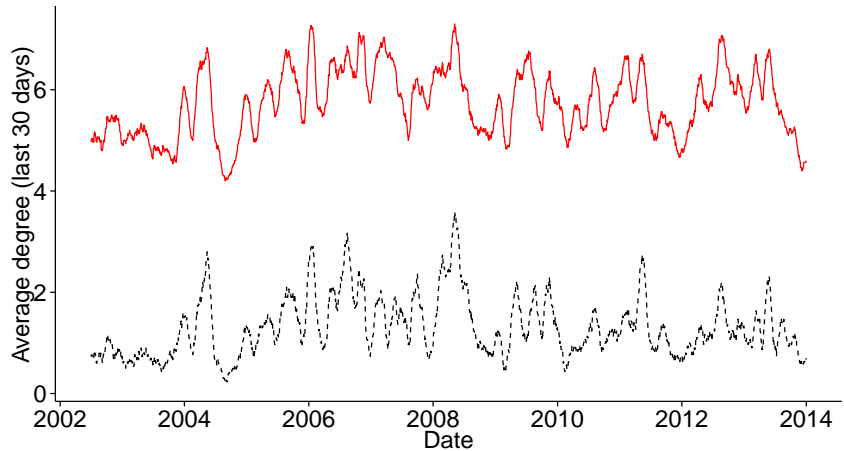}}
\vspace*{8pt}
\caption{The evolution of the maximum degree (solid line) and the difference between the maximum and the second largest degree (dashed line) within the minimally spanning trees based on the Randomized Dependence Coefficient. The values have been averaged over 30 consecutive networks before a given date to eliminate the noise. The evolution is not characterised by a particularly large period of interest, but it is worth noting that the networks change their structure periodically.\label{fig:fsdeg}}
\end{figure}

To show the detailed evolution of the networks themselves, in Fig.~\ref{fig:fsdeg} we present the evolution of the highest degree in the network in time, together with the evolution of the difference between the highest and second highest degree in the network in time.

\begin{figure}%
\centering
\subfloat[][]{%
\label{fig:dd-a}%
\includegraphics[width=0.5\textwidth]{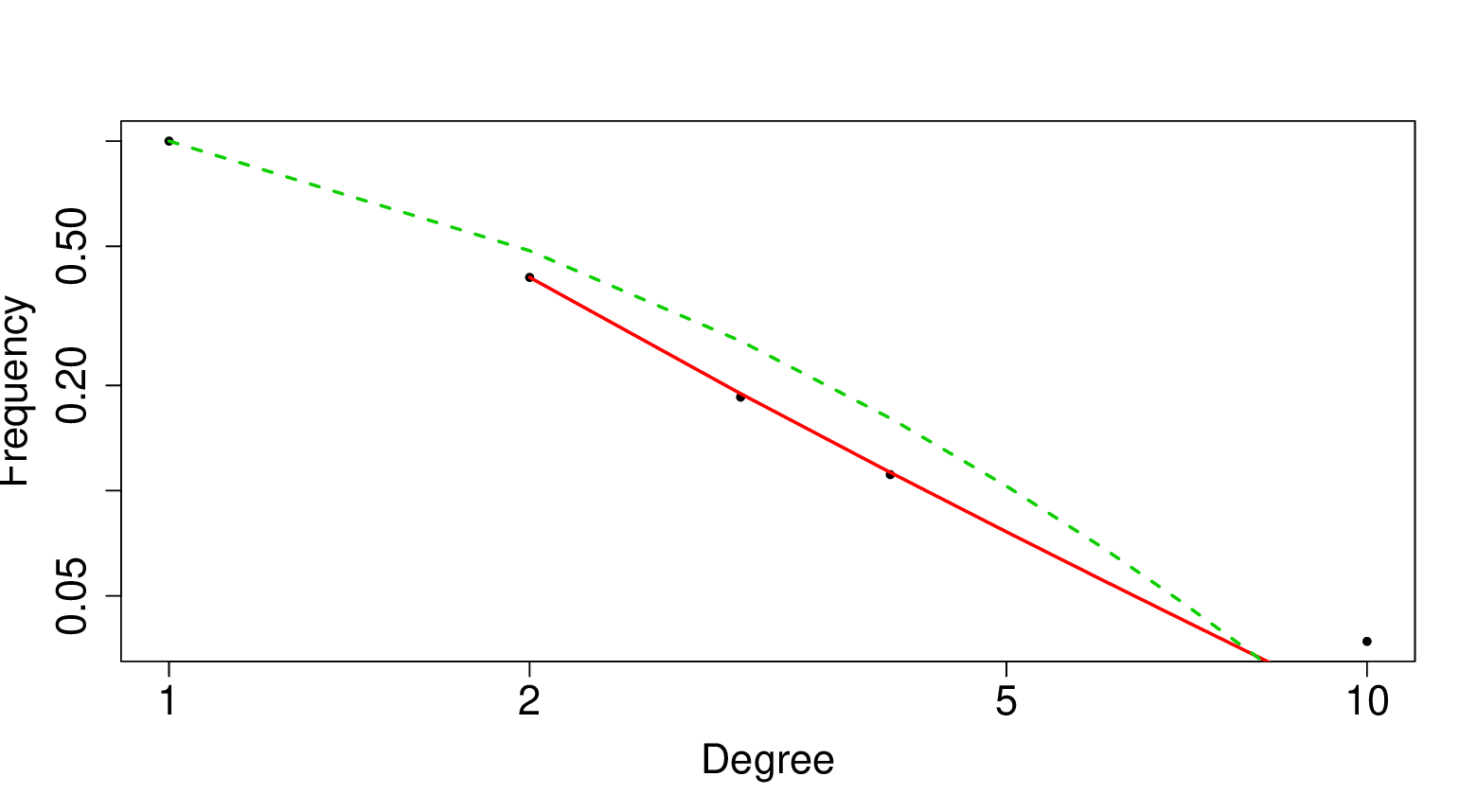}}%
\subfloat[][]{%
\label{fig:dd-b}%
\includegraphics[width=0.5\textwidth]{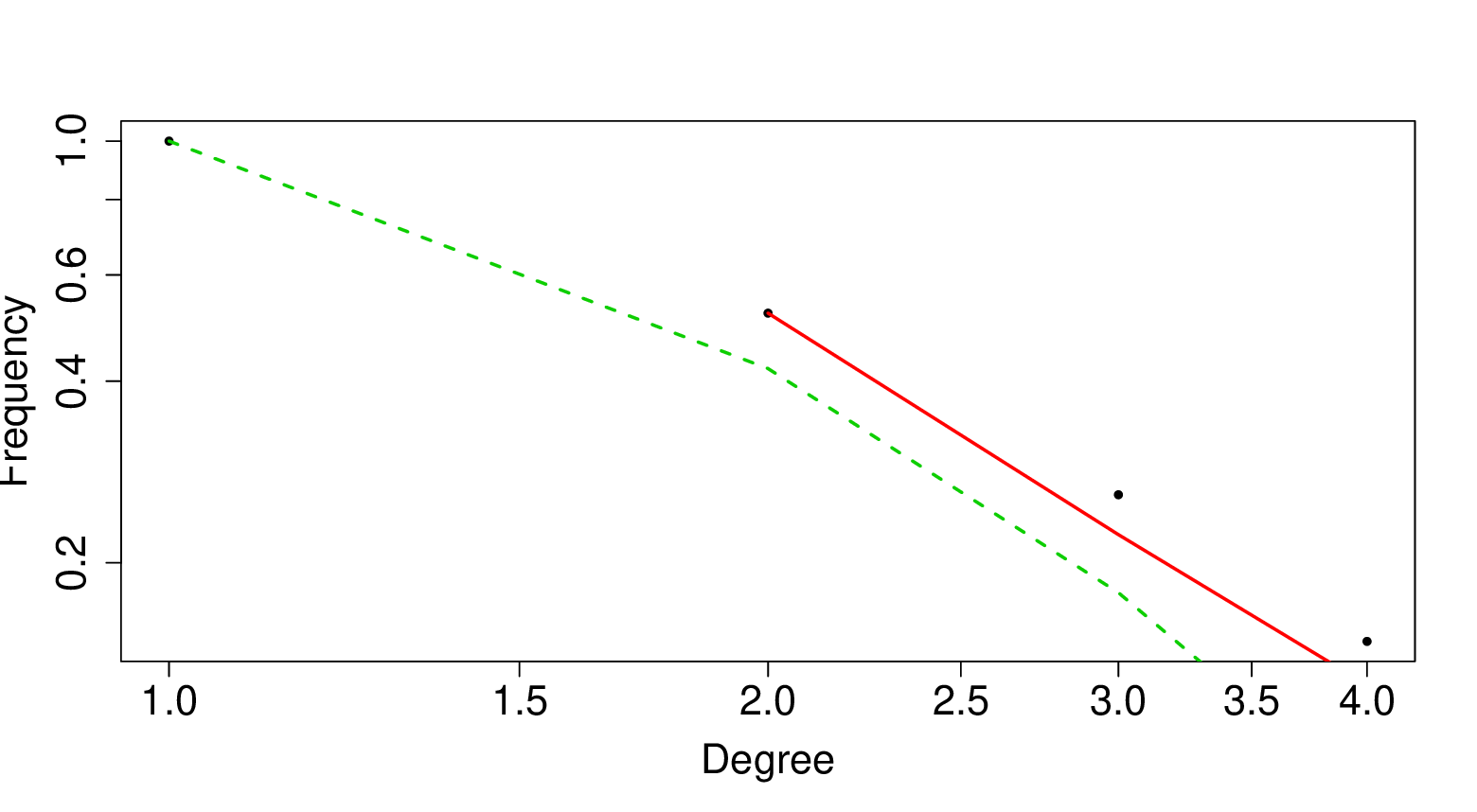}}%
\caption[]{Degree distributions, together with fitted power law (solid line) and log-normal (dashed line), of two particular networks based on the Randomized Dependence Coefficient:
\subref{fig:dd-a} a network where the difference between the highest and second highest degree achieves its maximum (2002-06-20);
\subref{fig:dd-b} a network where the difference between the highest and second highest degree achieves its minimum (2010-07-26).
Both networks show signs of fat tails, but technically a test for the presence of power law would fail in both cases.}%
\label{fig:dd}%
\end{figure}

To further present the different structure of the world currency network at different times, the degree distribution in two cases, one with the lowest difference between the maximum and the second largest degree (2002-06-20), and one with the largest difference between the maximum and the second largest degree (2010-07-26), have been presented in Fig.~\ref{fig:dd}

\section{Discussion}

First we turn to the question of the usefulness of the RDC in creating financial networks. As we do not have a theory of financial markets, we have no way of deciding which of the networks created is closer to the truth. We may find some intermediate measures to answer this question however. First, we know ex definitione that the RDC is a more general measure than Pearson's correlation coefficient, thus we expect our methodology to include all the relationships which are excluded in the standard methodology. Further, currency networks created with the standard methodology are structured based on geographical location, even though information about geography is not fed into the calculations, much as stock networks are structured based on economic sectors. This property is important, as it cannot be reproduced by simulating a market \cite{Bonanno:2000}. Thus we test whether this property is retained in the networks created with the RDC and compare the results with the ones obtained by using Pearson's correlation. As we have over 3000 networks we present these results as kernel distributions, in Fig.~\ref{fig:intra}, for networks created with a running window of length of 100 and 200 days. In the full graph we would find only 27.64\% of intracontinental links, but in the networks created with the RDC the average equals to 46.7\% and 47.5\% for 100 and 200 days long windows respectively. In comparison, the same values for the networks created with the Pearson's correlation coefficient equal to 48.3\% and 48\%. Standard deviation lies between $0.06$ and $0.07$ in all cases. There is another useful feature desired in financial networks, namely the existence of fat tails in the degree distributions. In Fig.~\ref{fig:dd} we present the degree distributions for networks created for two days, one in which the difference between highest and second highest degree within the network is maximal, and one in which it is minimal. We observe that while it is hard to pronounce these distributions to be formally following the power law, the are indeed characterised by fat tails. We find that in most studies analysing financial networks the insistence on power law is both theoretically not necessary, and practically unfounded. We thus find the obtained results satisfactory. Therefore, we are able to conclude that the networks created with the proposed methodology retain valuable properties even at short time horizons, and are potentially enhancing the analysis by the inclusion of the non-linear dependencies.

Having established the usefulness of the presented methodology we turn to the analysis of the obtained networks. In this, we concentrate on the analysis of the time evolution of degree distributions, as these are typically seen to be the most important characteristic in the study of networks, and particularly important in controllability of economic and social networks. The degree of a given currency within the world currency network approximates the importance of this currency, and its influence on the whole system. To get the big picture, and further analyse the differences between networks based on the RDC and Pearson's correlation we present the ranking of all studied currencies based on the average degree of the currency in all networks constructed when moving along the time series with the running window. The ranking is presented in Table~\ref{tab:ranks}. Please note that in parentheses we present the ranking as it would appear if we based the analysis on the Pearson's correlation. We find that our methodology correctly indentifies dollar as the most important currency. The rankings created with the two similarity measures are not very different. It is worth noting that our methodology shows the euro to be more important than the Danish krone, while analysis based Pearson's correlation shows the opposite. It is also noteworthy that these two currencies are closely tied together, and it may be difficult for such methodology to pick up on which currency is the leading one without further assumptions fed into the model. Nonetheless, it is worth noting that our methodology performs better in this respect. It is also interesting to note low positions of the Canadian dollar and the Brazilian real.

To show the evolution of the world currency market we present a shorter ranking, of only three main currencies, in all years between 2002 and 2013, thus showing the evolution of the main actors within the world market of currencies. The rankings, together with the degree of the currencies averaged over all networks created for a given year, are presented in Table~\ref{tab:yearly}. First, we note that the average degree of a leading currency varies between studied years. In particular, this measure has been much higher in the two years in which the financial crisis has been at its peak, that is the 2007 and 2008. This suggests that the dollar has monopolised a large part of the evolution of the world currency market in the times of crisis. This is important as it means the other strong currencies, such as the euro or the renminbi, are not able to successfully counteract the bad influence of the dollar in economically troubling times. This hints the need for the disentanglement of the world currency market, in order for it to become more resilient. Additionally, this analysis shows that while the euro has been the leading currency around the start of our analysis (2002--2003), it quickly lost its significance and dropped beyond the top ten in the rankings for the years of 2010--2011. It seems to be slowly regaining the position in the recent years however. This is troubling in the sense that the euro is supposed to be counteracting the power of the dollar, as hinted above. This role seems to lie more strongly on the Asian currencies in the recent years, notably on the Hong Kong dollar and the renminbi. Interestingly, the four mentioned currencies are the only ones which obtained the first place in the yearly rankings in our study. Thus we will look at the time evolution of their influence within the networks more closely.

In Fig.~\ref{fig:eurusd} we show the time evolution of the degree of the four mentioned currencies (averaged over 30 latest networks created before the given date to remove the statistical noise). Those are divided into two pairs: the Western currencies (EUR \& USD, Fig.~\ref{fig:eurusd-a}) and the Asian currencies (HKD \& CNY, Fig.~\ref{fig:eurusd-b}). The most interesting feature is the correlation between currencies from different groups and the lack of correlation between currencies within the groups. There is a negative correlation between degrees of Western and Asian currencies of $-0.337$ for CNY/USD, $-0.204$ for CNY/EUR, $-0.264$ for USD/HKD, and $-0.371$ for EUR/HKD. Pearson's correlation equals $0$ for both USD/EUR and CNY/HKD pairs. This means that the balance of power is decided between the Western and the Asian currencies, depending on the situation in the world. In particular the Western currencies were much more important to the evolution of the global system in the years of the crisis (2007--2008), while the Asian currencies were more important in the years before and after this. It is worth noting that the importance of all four currencies seem to be changing periodically, and it may potentially be modelled and predicted in certain situations using log-periodic approach \cite{Bartolozzi:2005}. Regarding the current state of the market, it is worth noting that at the start of the year 2014 the importance of Asian currencies is rising again.

Finally, we analyse the concentration of power (in the form of high degree) within the world currency network. In Fig.~\ref{fig:fsdeg} we present the highest degree averaged over the last 30 networks created by a given date, and the difference between the highest and second highest degree averaged over the last 30 networks created by a given date. Both show the mentioned concentration. Unlike the studies looking at the stock markets \cite{Sienkiewicz:2013,Wilinski:2013,Kozlowska:2014} we do not find dramatic phase transitions. Nonetheless, the world currency system seems to have been more steady around the years 2002--2003, while in the later years it seems to be changing structure periodically. In particular we find the largest concentration of power in the USD in the early 2008, which we have seen above. Importantly, this may serve as a measure of risk for companies and other entities whose interest are spread across many currencies. When these measures are high, the world currency network is concentrated and thus synchronised, increasing the risk of even the portfolios diversified around various currencies. On the other hand, the situation such as at the end of the year 2013, where the highest degree in the world currency network is at the lowest value in our study (around 4), is more beneficial from the perspective of global currency risk. Nonetheless, a more fitting analysis of risk must be crafted to a specific institution, in particular should use currencies denominated in the home currency of this institution and not in silver. Otherwise the methodology will remain the same.

\section{Conclusions}

In this study we have presented a method for creating financial networks which can account for non-linear dependencies, does not require discretisation of the data, and works well even for short time horizons. We have shown that this method creates networks with fat tails, and reconstructs the geographical structure of the world currency network from the rate changes as well as the standard method based on Pearson's linear correlation coefficient. We propose this method to be better suited for most analyses of financial networks than the standard methodology, particularly for short time horizons. For studies with large samples methods bases on information--theoretic measures such as mutual information may be performing better however \cite{Fiedor:2014a}. We have applied this procedure to dataset of currency exchange rates for 27 major world currencies in relation to silver, and created networks with a running window of width equal to 100 days. Thus we were able to comment on the time evolution of the world currency network between the year 2000 and 2013, as well as the changing influence of various currencies. Most notably we have seen the fading importance of the euro since 2004. Importantly, the world currency system seems not to be very resilient with regard to shocks, and has been very synchronised in following the dollar in the years 2007 and 2008. We have also found the interplay between two major Western (USD \& EUR) and two major Asian currencies (HKD \& CNY), which exchange their significance to the world market periodically. These shifts of power and concentration within the world currency networks may potentially be modelled and predicted using log-periodic approach, which should be the subject of further studies. A more detailed study should be performed to analyse the performance of the Randomized Dependence Coefficient in the financial market data based on other non-linear association patterns. Further studies should also be performed to investigate the time evolution of other financial networks, in particular various stock markets.

\appendix

\section{Currencies used in the study}

\begin{table}[ht]
\tbl{Currencies used in this study.}
{\begin{tabular}{llllll}
\toprule
Code & Name & Code & Name & Code & Name \\ \colrule
ARS & Argentine Peso & EUR & Euro & PKR & Pakistan Rupee \\
AUD & Australian Dollar & FJD & Fiji Dollar & PHP & Philippine Peso \\ 
BRL & Brazilian Real & HKD & Hong Kong Dollar & RUB & Russian Ruble \\
GBP & Pound Sterling & INR & Indian Rupee & SEK & Swedish Krona \\
CAD & Canadian Dollar & ILS & New Israeli Shekel & THB & Thai Baht \\
CHF & Swiss Franc & MYR & Malaysian Ringgit & TRY & New Turkish Lira \\
CLP & Chilean Peso & MXN & Mexican Peso & TWD & New Taiwan Dollar \\
CNY & Yuan Renminbi & NZD & New Zealand Dollar & USD & US Dollar \\
DKK & Danish Krone & NOK & Norwegian Krone & ZAR & South African Rand \\ \botrule
\end{tabular}
\label{tab:ccy}}
\end{table}

\end{document}